\documentclass[10pt,a4paper,twoside]{article}
\usepackage{epsfig}
\usepackage{baltlat6}
\usepackage{array}
\usepackage{here}
\usepackage{changes}
\begin{document}
\ \
\vspace{0.5mm}
\setcounter{page}{1}

\titlehead{Baltic Astronomy, vol.\,xx, xxx--xxx, 2016}

\titleb{HIDDEN POPULATION OF ALGOLS}

\begin{authorl}
\authorb{O. Malkov}{1},
\authorb{D. Kovaleva}{1},
\authorb{L. Yungelson}{1},
\authorb{E. Avvakumova}{2},
\authorb{D. Chulkov}{1},
\authorb{O.~Dluzhnevskaya}{1},
\authorb{A. Kniazev}{3,4,5,6}
\end{authorl}

\begin{addressl}
\addressb{1}{Institute of Astronomy, Russian Acad. Sci.,
48 Pyatnitskaya St., Moscow 119017, Russia;
malkov@inasan.ru}
\addressb{2}{Kourovka Astronomical Observatory, Institute of Natural Science,
B.N.~Yeltsin Ural Federal University, 19 Mira St., Yekaterinburg
620000, Russia}
\addressb{3}{South African Astronomical Observatory, PO Box 9, 7935 Observatory,
South Africa}
\addressb{4}{Southern African Large Telescope Foundation, PO Box 9, 7935 Observatory,
Cape Town, South Africa}
\addressb{5}{Sternberg Astronomical Institute, Moscow State University,
13 Universitetskij Prosp., Moscow 119992, Russia}
\addressb{6}{Special Astrophysical Observatory, Russian Acad. Sci., Nizhnij Arkhyz,
Karachai-Cherkessian Republic 369167, Russia}

\end{addressl}

\submitb{Received: 2016 xxx xx; accepted: 20xx xxx xx}

\begin{summary}
We present results of Monte Carlo simulation aiming at the
estimate of the frequency of semi-detached Algol-type binaries
among the stars observed as single ones. When account is made for
various detection biases (mostly due to inclination of orbits),
the fraction of Algols among Galactic disk stars appears to be
0.1--0.2\%. However, this number should be regarded as a lower
limit only, since there are still unaccounted selection effects
and other types of photometrically unresolved binaries. Hidden
binarity appears to be an important phenomenon that should be
taken into account when considering stellar statistics and
construction of  fundamental relations between stellar parameters.
\end{summary}


\resthead{Hidden population of Algols}
{O.~Malkov et al.}

\sectionb{1}{INTRODUCTION}

Binary stars can be detected by various techniques. Components of
the closest pairs can be resolved, so that the stars are observed
as visual or interferometric binaries. For more distant (and,
consequently, photometrically unresolved) binaries, a successful
combination of the inclination and size of the orbit, as well as
of components' parameters, can lead to observed eclipses and/or
Doppler shift of spectral lines. Close binaries in late stages of
their evolution can demonstrate variability in X-ray or radio
emission.

However, at least a part of binary systems remain undiscovered by
modern techniques. In particular, a close binary, observed as an
eclipsing variable, would not be detected were its orbit
inclination less than 30--40$^\circ$, as we show below. Such
``pole-on'' binaries look like single stars. As a result, the
single  star statistics is biased, and we should take this into
account when constructing calibrating relations like the
mass-luminosity relation (Malkov 2007), compiling the luminosity
function (discussed in Piskunov \& Malkov 1991 and Malkov et al.
1998 for main-sequence and pre-main-sequence stars, respectively),
or estimating the local missing mass (Malkov 1994).

The principal goal of this work is to compare the numbers of stars
classified as Algol-type eclipsing binaries (hereafter Algols)
based on their light curves and those semi-detached systems with
similar components where eclipses are not observed. In the present
study, we limit ourselves to Algols, as they are one of the most
representative types of eclipsing binaries. Stars of other types
will be examined in further studies.

\sectionb{2}{CLASSIFICATION OF ALGOLS}
\label{sec:classification}

A classification scheme for  semi-detached systems was proposed by
Popper (1980), who divided them into three groups: (i) the more
massive systems in which the hotter component is an early B-type
star and the cooler one, a star of type B or early A; (ii) the
more ``typical'' Algol systems of lower mass in which the more
massive component has a spectral type in the range from mid-B to
early F and the companion is of type F or later; (iii) later-type
subgiant and giant semi-detached systems.
In the present study, we deal with group (ii) stars.

One of the most comprehensive sources of data on Algols is the
Catalogue of (411) Algol type binary stars by Budding et al.
(2004). It contains, in particular, data on physical parameters of
the components (mass, luminosity, temperature, radius, etc.), when
known. Analysis of these data together with those from the Surkova
\& Svechnikov (2004) catalogue of Algols shows that, in the
majority (about 70\%) of Algols, the accretor is hotter, smaller,
more massive and more luminous than the donor. In the remaining
30\% of objects, the accretor is hotter, {\it larger}, more
massive and more luminous than the donor. Hereafter we call such
stars ``regular'' and ``inverse'' Algols, respectively.

It should be noted that, according to Budding et al. (2004) and
Surkova \& Svechnikov (2004) data, there are a few (about a dozen)
so-called ``rare'' Algols, where the accretor is hotter, smaller,
more massive and {\it less luminous} than the donor. Examples are
HH~Car and KU~Cyg. Finally, there is an extremely small class of
the (so-called ``marginal'') Algols, where the accretor is hotter,
smaller, {\it less massive} and more luminous than the donor.
SS~Cam belongs to that class, and about five more systems are
considered to be candidates to ``marginal'' Algols.

Assuming similar conditions for formation of all Algols, it can be
supposed that ``marginal'', ``rare'', ``regular'' and ``inverse''
Algols represent four consequent stages of evolution of a
semi-detached binary. Note that, in all cases, the accretor is
hotter than the donor.

\sectionb{3}{MONTE CARLO SIMULATIONS}

We generate a set of objects with primary constrained pairing (here we use the
classification scheme terms of pairing scenaria proposed by
Kouwenhoven et al. 2008) of objects. The mass $m_a$ of the primary
(hotter, more massive accretor)  is drawn from a pre-assumed mass
distribution $f_a(m)$, and the mass ratio $q$ is drawn from a mass
ratio distribution $\eta(q)$. Finally, the donor mass $m_d = qm_a$
is calculated. We use $f_a(m) \propto m^{-2.5}$ for $m_a$ from 1
to 5 $m_\odot$, adopted from Eretnova and Svechnikov (1994, their
Fig.~1b for low-mass short-period semi-detached systems (so-called
R~CMa systems) and ``typical'' semi-detached systems), and
$\eta(q) \propto q^{-2.8}$ for the mass ratio range from 0.10 to
0.75 (adopted from Svechnikov et al. 1989, their Fig.~20a).

According to evolutionary considerations, we discard pairs where
$m_d \le 0.15 m_\odot$ or $m_d+m_a>6.25 m_\odot$.
Here, the lower limit for $m_d$ is the minimum mass of a He
white dwarf, while the upper limit for the total mass is defined
by adopted limit of $m_a\leq5$.

It was shown by Popova et al. (1982) that the distribution of
binaries over semi-major axes of orbits $a$ was satisfactorily
described by the function $dN \sim d\log a$ for $10 R_\odot \le a
\le 10^6 R_\odot$. Eretnova and Svechnikov (1994, their Fig.~3b,
upper panel) confirm this distribution, and we use it over the
$\log a$ range from 0.7 to 1.5. All systems are assumed to be
circularized.

The orbits were assumed to be equiprobably oriented, resulting in
a distribution of inclinations proportional to $\sin i$. Systems
have a homogeneous spatial distribution. To estimate interstellar
extinction, we use the standard cosecant law (Parenago 1940) with
the coefficients $a_0=0.0016$ mag/pc and $\beta=114$ pc, found by
Sharov (1964).

Thus, we varied $m_a$, $q$, all orbital elements, and the distance
to the system according to the above-mentioned functions.

The presence and depth of an eclipse in a system depend on
components' radii and temperatures and on the orbital inclination.
To estimate effective temperatures of donors and accretors, we
used the $T_{\rm eff}$ -- mass relations from Svechnikov et al.
(1989). Radii of donors and accretors were estimated using
relations from Svechnikov et~al. (1989) and Gorda \& Svechnikov
(1998), respectively. Accretors were assumed to satisfy the
main-sequence mass-radius relation, and donors were assumed to
fill their Roche lobes. The Roche lobe radius was calculated
according to Eggleton (1983).

\begin{figure}[!tH]
\vbox{
\centerline{\psfig{figure=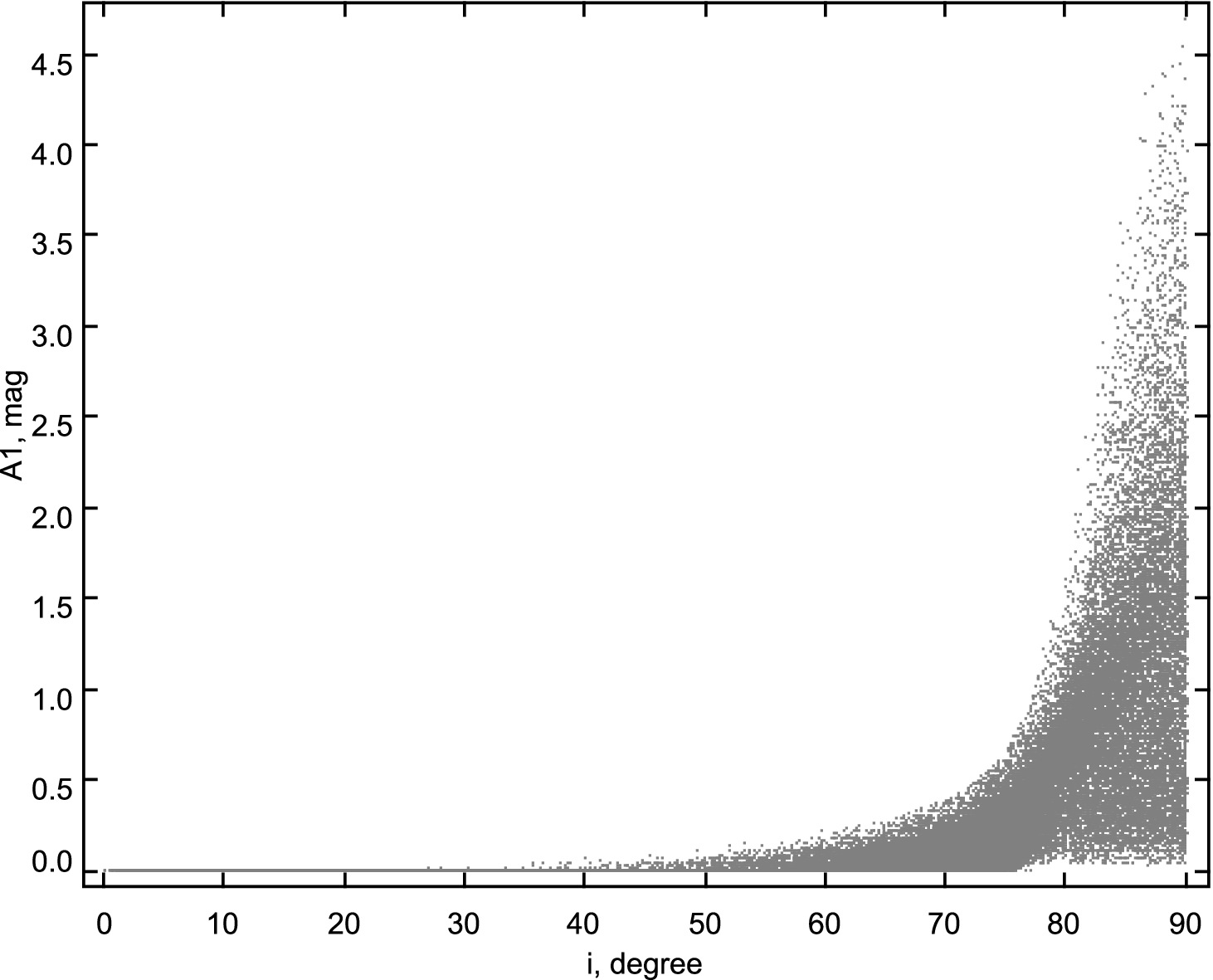,width=100mm,angle=0,clip=}}}
\vspace{1mm}
\captionb{1}
{Depth of primary minimum vs. orbital inclination for simulated binaries}
\end{figure}

We collected statistics of eclipsing and non-eclipsing simulated
systems separately. If a given system happens to be eclipsing, we
calculate its brightness in maximum, depth of primary minimum, and
orbital period. Dependence of the primary minimum depth $A_1$ on
the orbital inclination $i$ for simulated systems is shown in
Fig.~1. It can be seen that $A_1$ can reach 4.5 mag (systems with
maximum $T_{\rm eff}$ difference of components demonstrate values
that large) and that ``extremely edge-on'' systems ($i>75^\circ$)
cannot produce very shallow ($A_1<0^m.01$) minima.

\sectionb{4}{OBSERVATIONAL DATA}

To verify our procedure, we have compared characteristics of the
simulated eclipsing binaries to those of observed Algols. The
latter ones were drawn from the Catalogue of Eclipsing Variables,
CEV, compiled by Malkov et al. (2006) mainly from the General
Catalogue of Variable Stars, GCVS (Samus et al. 2013), upgraded by
Avvakumova et al. (2013), and uploaded into the Binary stars
database, BDB (Kaygorodov et al. 2012, Kovaleva et al. 2015).
The results of our simulations were tested against a CEV-based
sample of  415 Algols with original classification and 1726
systems classified as Algols using the procedure suggested by
Malkov et al. (2007) and substantially modified by Avvakumova \&
Malkov (2014), altogether 2141 systems.

\begin{figure}[!tH]
\vbox{
\centerline{\psfig{figure=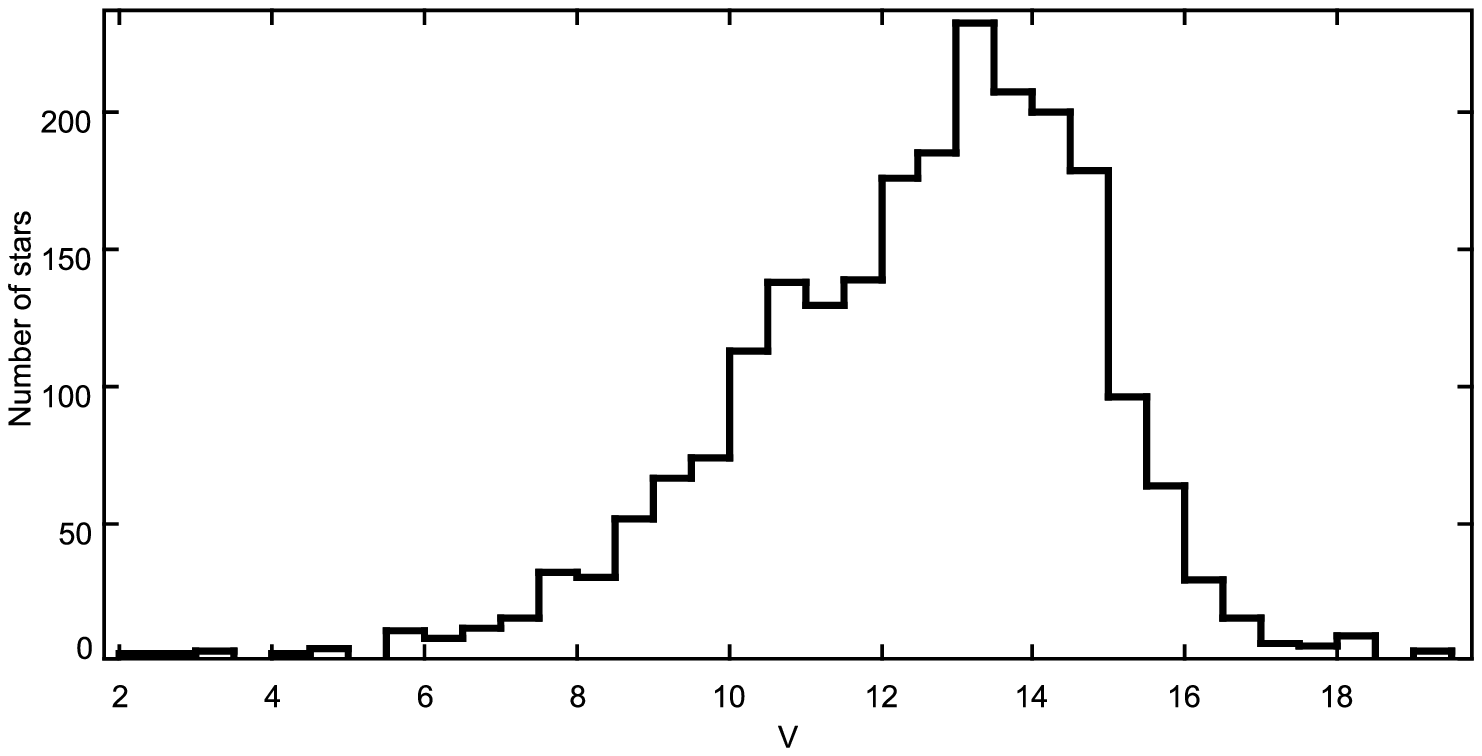,width=100mm,angle=0,clip=}}}
\vspace{1mm} \captionb{2} {Distribution of Algols, catalogued in
CEV, over brightness.}
\end{figure}

Figure~2 illustrates the brightness (magnitude in maximum)
distribution of catalogued Algols. One can see that the sample can
be considered complete down to $V=10.^m5$. This value will be used
to correct the simulated sample for incompleteness.

\begin{figure}[!tH]
\vbox{
\centerline{\psfig{figure=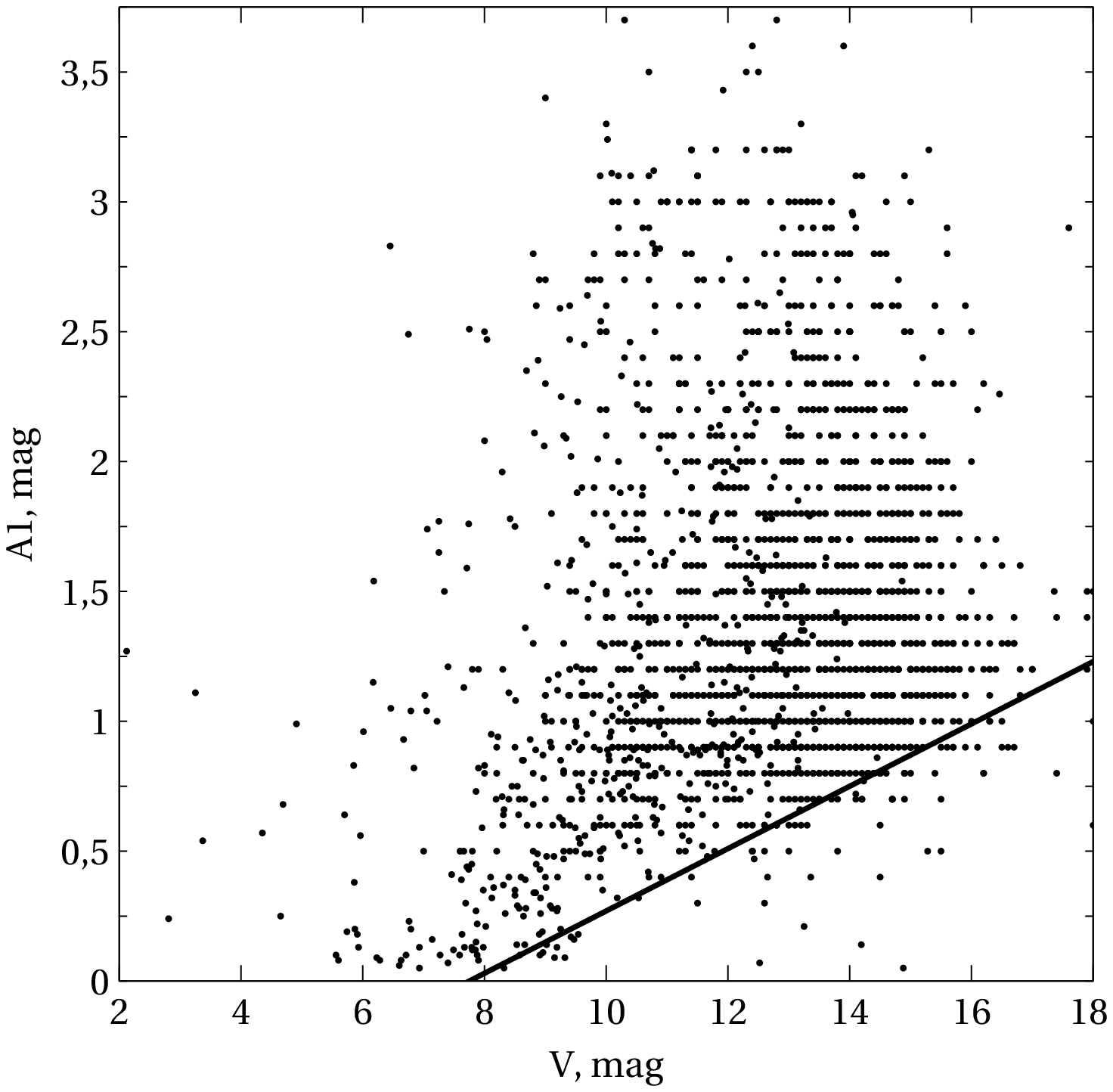,width=100mm,angle=0,clip=}}}
\vspace{1mm}
\captionb{3}
{Depth of primary minimum vs. brightness of Algols, catalogued in CEV.
An ``avoidance triangle'' in the lower right corner is discussed in the text.}
\end{figure}

Obviously, the simulated sample should be corrected for some
selection effects. The depth of the primary minimum $A_1$ versus
maximum system brightness is plotted in Fig.~3. An ``avoidance
triangle'' in the lower right corner reflects the difficulties in
observations of shallow minima for faint stars. Consequently, we
have discarded all stars with $A_1 < 0.12 \cdot V - 0.93$ (the
solid line in Fig.~3) from the final statistics. Also, one can
note a lack of stars in the upper right corner resulting from the
absence of faint, high-amplitude systems in observational
statistics: in their minimum brightness, they are too faint to be
detected, classified, and included in the catalogues.

Besides the stars fainter than $V=10.^m5$ and stars that fall in
the ``avoidance triangle'' (see the previous paragraph), we have
also removed stars with $A_1<0^m.05$, as there are no such
``extremely shallow minimum'' Algols in the CEV catalogue.

\begin{figure}[!tH]
\vbox{
\centerline{\psfig{figure=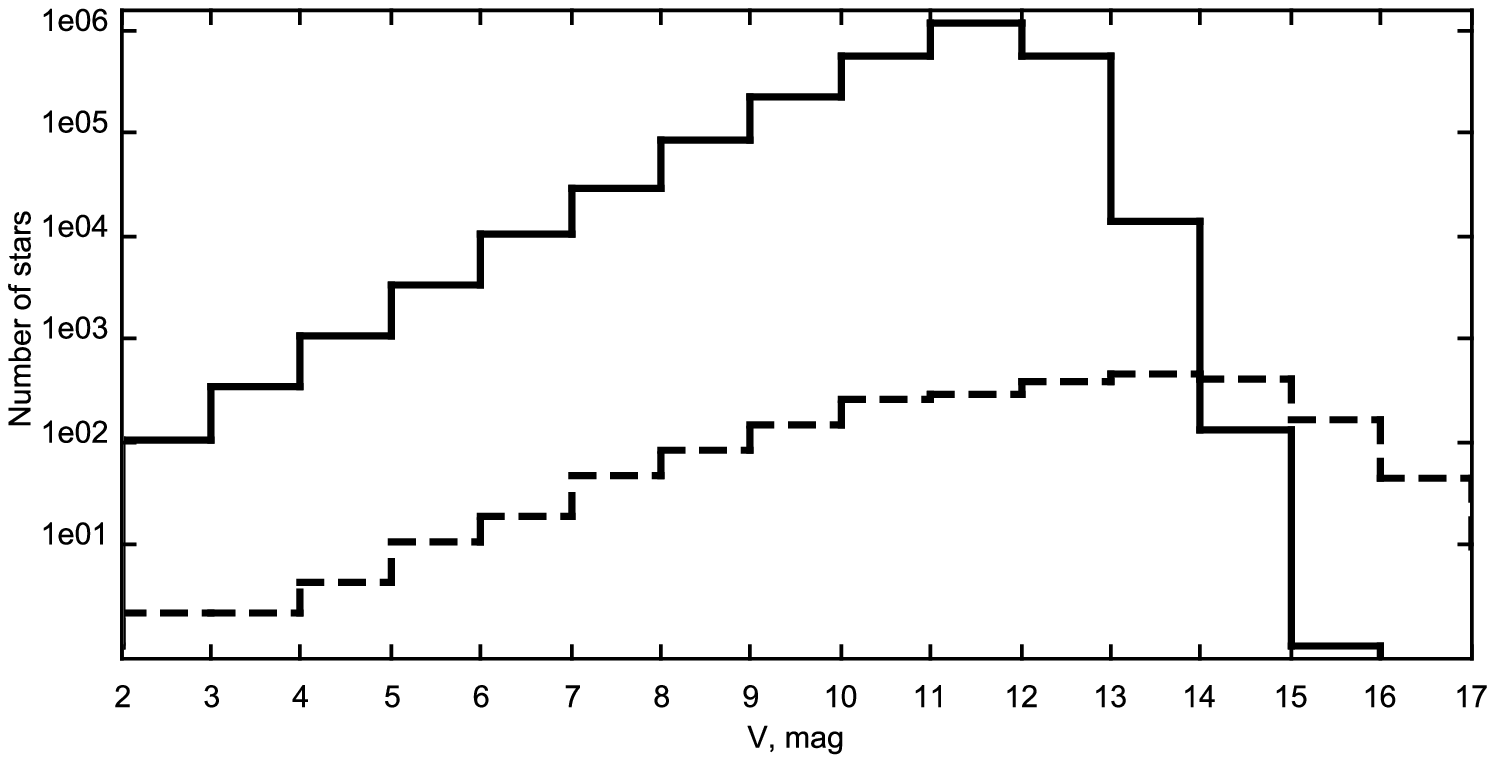,width=100mm,angle=0,clip=}}}
\vspace{1mm} \captionb{4} {Distribution of Tycho-2 catalogue stars
(solid line) and Algols from the CEV catalogue (dashed
line) over brightness.}
\end{figure}

Distribution of the catalogued Algols over the brightness is
compared with that for the Tycho-2 Catalogue (H\o g et al. 2000)
stars in Fig.~4.

\sectionb{4}{COMPARISON OF SIMULATED AND OBSERVATIONAL DATA}

\begin{figure}[!tH]
\vbox{
\centerline{\psfig{figure=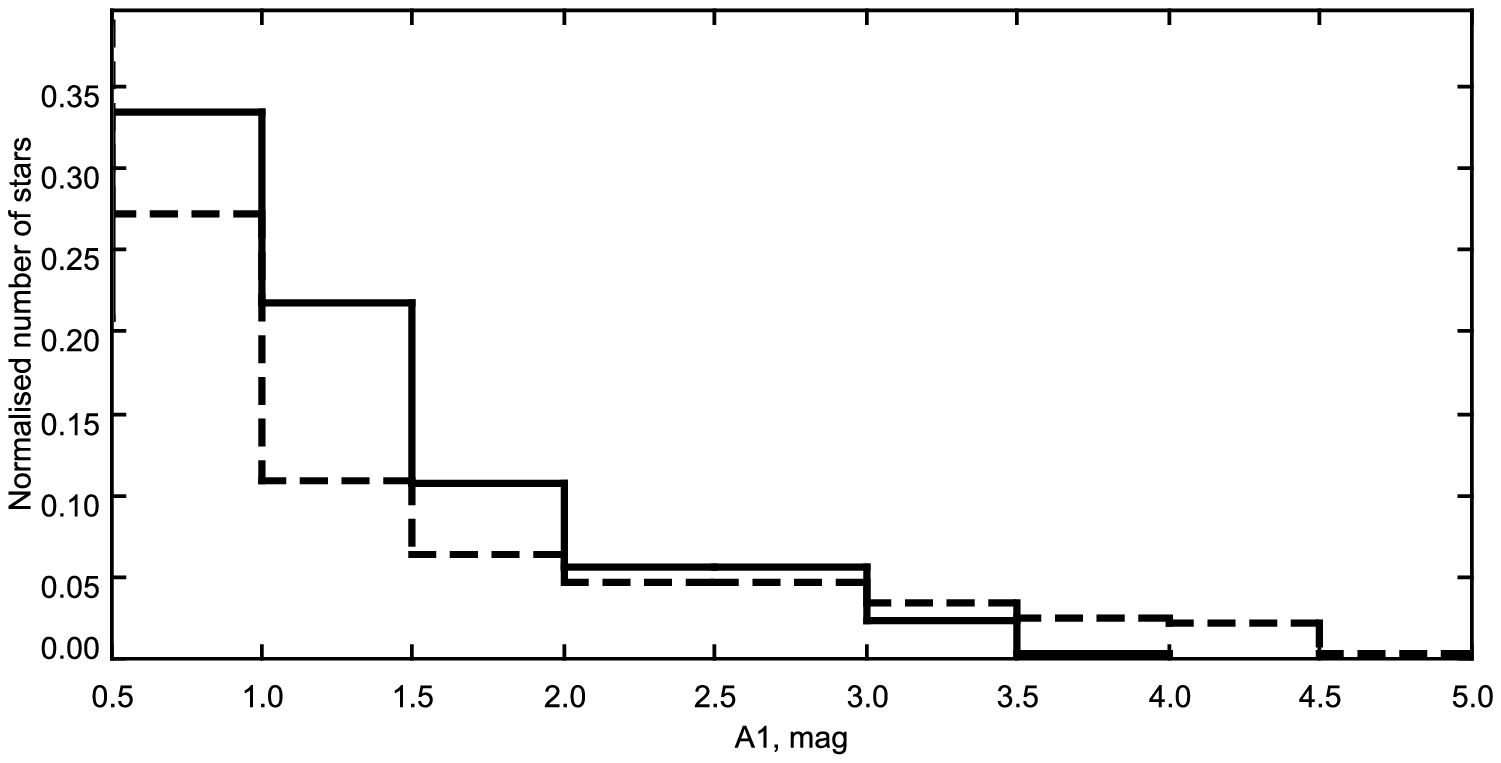,width=100mm,angle=0,clip=}}}
\vspace{1mm}
\captionb{5}
{Catalogued (solid line) and simulated (dashed line) Algols:
depth of primary minimum distribution.}
\end{figure}

\begin{figure}[!tH]
\vbox{
\centerline{\psfig{figure=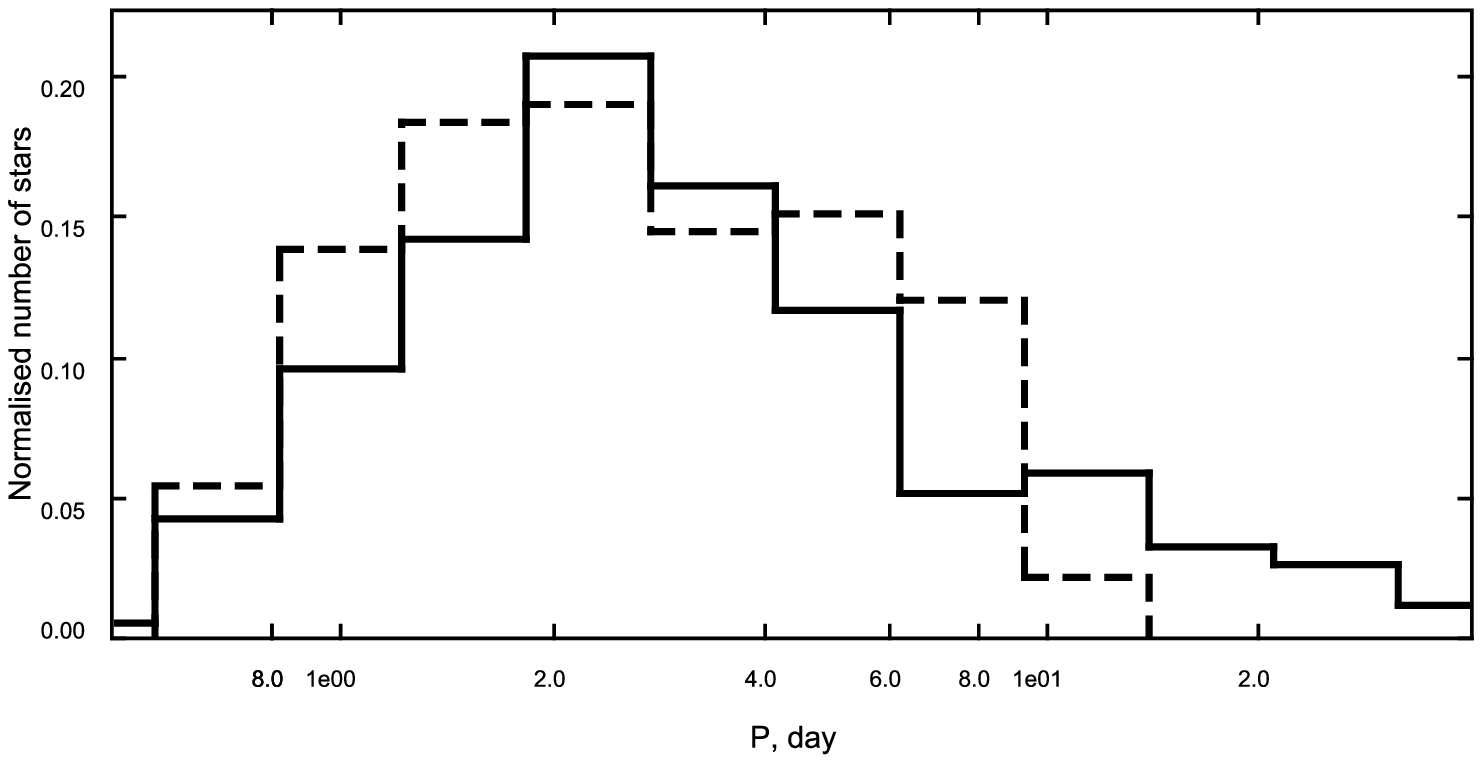,width=100mm,angle=0,clip=}}}
\vspace{1mm} \captionb{6} {Catalogued (solid line) and
simulated (dashed line) Algols: the orbital period
distribution.}
\end{figure}

Distributions of simulated and catalogued Algols over the depth of
primary minima (Fig.~5) and the orbital periods (Fig.~6)
demonstrate a satisfactory agreement between calculations and
observations. In particular, we have managed to reproduce a
quasi-triangle period distribution with a maximum at $\sim
2$~days. However, a few longest-period (periods longer than $\sim
15$ days) systems could not be simulated, probably due to too
small upper limit for the semi-major axis.

Our simulation procedure produces both ``regular'' and ``inverse''
Algols (see Section~2). The resulting fraction of simulated
``inverse'' eclipsing binaries among all, ``regular'' and
``inverse'', systems is 0.22, which is reasonably close to the
observational value of 0.3.

\sectionb{5}{RESULTS AND DISCUSSION}

\begin{figure}[!tH]
\vbox{
\centerline{\psfig{figure=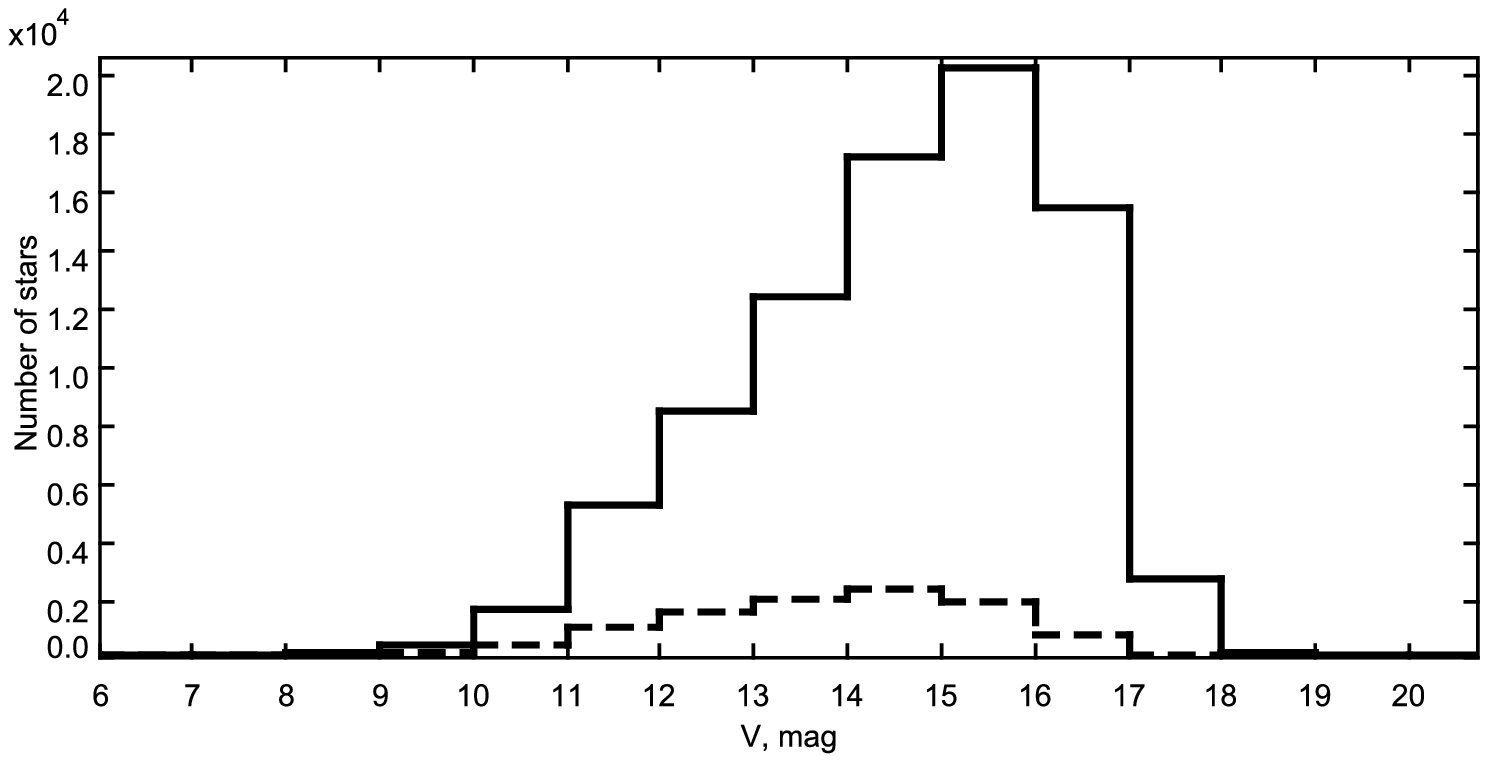,width=100mm,angle=0,clip=}}}
\vspace{1mm} \captionb{7} {Distribution of simulated systems over
brightness: all systems (solid line) and systems with
eclipses (dashed line).}
\end{figure}

After verification of the procedure, we compare, for our
simulation, the number of Algols with eclipses and number of
systems where eclipses are not observed. Distributions of {\it
all} simulated systems and systems with eclipses over their
brightness are shown in Fig.~7.

To assess the extent to which ``pole-on'' Algols can distort the
observational statistics of single stars, one should compare, for
a given visual magnitude, the number of all observed stars and the
number of observed Algols, corrected for systems where eclipses
are not observed. This can be estimated from Figs.~4 and~7. For
$V=8^m$, for instance, per 1000 Tycho-2 stars, one Algol is
observed (and catalogued in CEV), and one more is hidden due to
the ``pole-on'' orientation. Corresponding estimates for
$V=10.^m5$ give a similar result: every 2500 Tycho-2 stars contain
one observed and 4.3 hidden Algols.

So, as our analysis shows, some 0.1 to 0.2\% of observed stars are,
in fact, semi-detached binaries. Can we ignore this effect
when making statistics of single stars? Probably not, due to
the following reasons.

\begin{itemize}

\item The effect is not evenly distributed in the HRD. The
majority of Algols, when observed as a single star, are classified
as MS stars. On the other hand, Tycho-2 catalogue stars, used for
comparison, are of {\it all} luminosity classes, and their
additional selection should be performed to make our estimates
more correct.

\begin{figure}[!tH]
\vbox{
\centerline{\psfig{figure=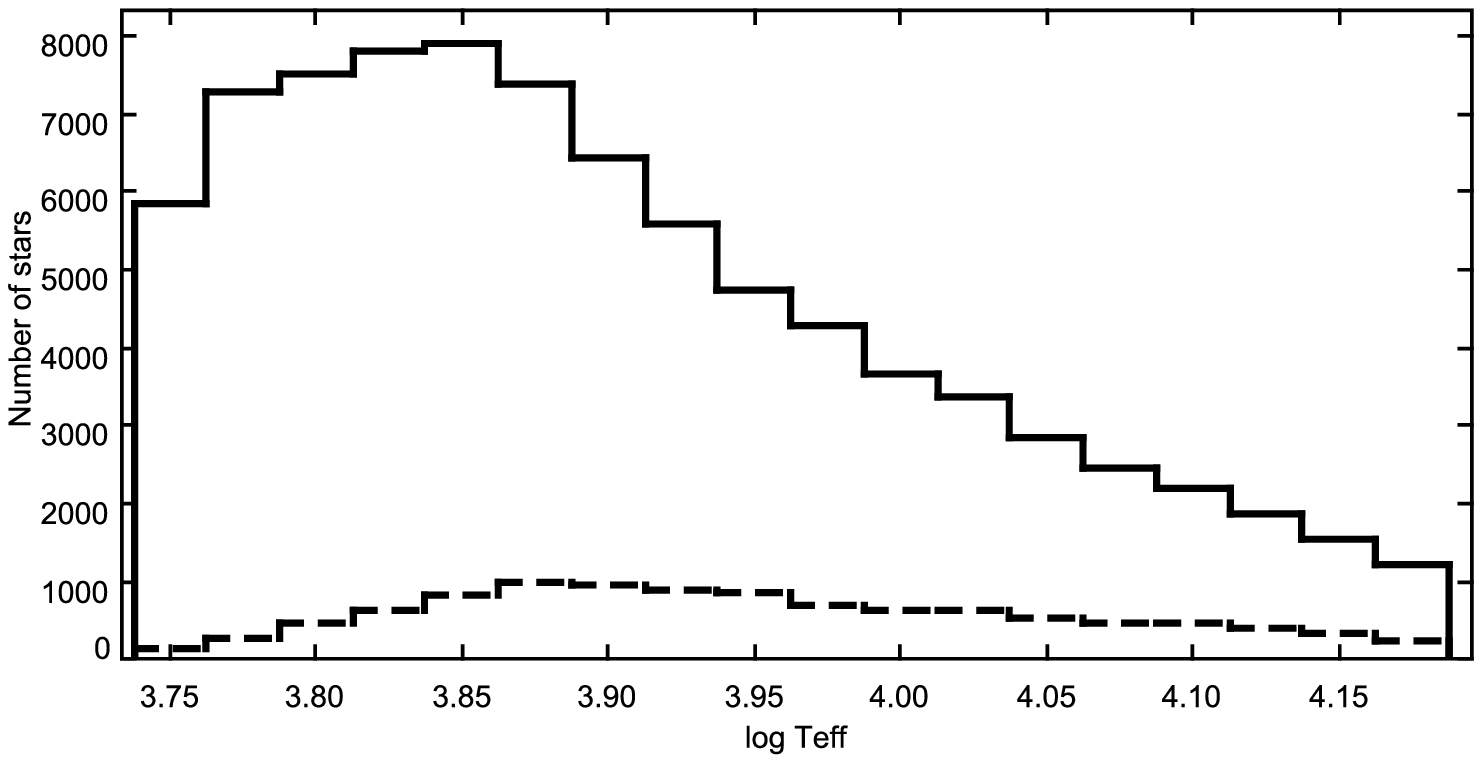,width=100mm,angle=0,clip=}}}
\vspace{1mm} \captionb{8} {Distribution of simulated systems over
the accretor's effective temperature: all systems (solid
line) and systems with eclipses (dashed line).}
\end{figure}

\item Further, our results are averaged over B to G spectral types,
but, as Fig.~8 shows, the effect is stronger for cool stars.
Consequently, statistics of F-type stars is distorted stronger
than our averaged estimates predict.

\item We have studied only one (though one of the most numerous)
type of eclipsing binaries. Other representative types (detached
main-sequence stars, contact W~UMa stars, etc.) also contribute to
the problem.

\item In the present study, we do not deal with systems undergoing
the second mass exchange. Under certain conditions, a compact
object, accreting the matter from its donor, may not produce X-ray
or UV radiation and, consequently, will be observed as a single
star. They appear to be not very important, but it is probably
advisable to estimate a number of such systems and their influence
on MS-star statistics.

\item Here we assume that if a system can demonstrate eclipses
with minimum depth $A_1>0^m.05$, it is discovered as a binary with
probability $P=1$. Actually, this is not the case unless we deal
with results of automatic surveys for variable stars, microlensing
events, or exoplanets. Usually such surveys cover only small area
on the sky.

\item In the comparison of our model to catalogued data, we were
restricted by the CEV completeness limit for Algols. However, as
can be seen in Fig.~7, the fainter objects we consider, the larger
the fraction of hidden Algols is, i.e. the stronger is the effect.

\item We did not account for other observational biases, namely,
some long period systems are missed in the model.
However, this effect cannot be very important due to relatively small
number of long-period systems (see, e.g., Fig.~6).

\end{itemize}

All of the listed reasons magnify the effect described above and
increase, to varying degrees, relative amount of photometrically
unresolved binaries among observed ``single'' stars. We suppose
that estimates made in this paper should be increased by an order
of magnitude and will reach 1 to 2\%, but this is a subject of a
future study.

\thanks{
The work was partly supported by the Presidium of Russian Academy
of Sciences program ``Leading Scientific Schools Support''
9951.2016.2, by the Russian Foundation for Basic Research grants
15-02-04053 and 16-07-01162,
by the Government of the Russian Federation, project no. 02.A03.21.0006,
and by the Ministry of Education and Science of the Russian Federation
(the basic part of the state assignment, registration number 01201465056).
A.\,K. acknowledges support from the National Research Foundation
of South Africa and from the Russian Science Foundation (project
No. 14-50-00043). This research has made use of the VizieR
catalogue access tool, CDS, Strasbourg, France, and NASA's
Astrophysics Data System Bibliographic Services. }

\References

\refb Avvakumova E. A., Malkov O. Yu. 2014, MNRAS, 444, 1982

\refb Avvakumova E.~A., Malkov O.~Yu., Kniazev A.~Yu. 2013, AN,
334, 859

\refb Budding E., Erdem A., Cicek C. et al. 2004,
A\&A, 417, 263

\refb Eggleton P. P. 1983, ApJ, 268, 368

\refb Eretnova O. V., Svechnikov M. A. 1994, Astronomy Reports,
38, 483

\refb Gorda S. Yu., Svechnikov M. A. 1998, Astronomy Reports, 42,
793

\refb H\o g E., Fabricius C., Makarov V. V. et al. 2000, A\&A,
355, L27

\refb Kaygorodov P., Debray B., Kolesnikov N. et al.
2012, Baltic Astronomy, 21, 309

\refb Kovaleva D. A., Kaygorodov P. V., Malkov O. Yu. et al.
2015, Astronomy \& Computing, 11, 119

\refb Kouwenhoven M. B. N., Brown A. G. A., Goodwin S. P. et al.
2008, AN, 329, 1

\refb Malkov O. Yu. 1994, Astrophysics, 37, 256

\refb Malkov O. Yu. 2007, MNRAS, 382, 1073

\refb Malkov O., Piskunov A., Zinnecker H. 1998, A\&A, 338, 452

\refb Malkov O.~Yu., Oblak~E., Snegireva E.~A., Torra~J. 2006,
A\&A, 446, 785

\refb Malkov O. Yu., Oblak E., Avvakumova E. A., Torra J. 2007,
A\&A, 465, 549

\refb Parenago P. P. 1940, AZh, 17, 3

\refb Piskunov A. E., Malkov O. Yu. 1991, A\&A, 247, 87

\refb Popova E. I., Tutukov A. V., Yungelson L. R. 1982, Ap\&SS,
88, 55

\refb Popper D. M. 1980, ARAA, 18, 115

\refb Samus N. N., Durlevich O.~V., Kazarovets E.~V. et al. 2013,
VizieR On-line Data Catalog: B/gcvs

\refb Sharov A. S. 1964, SvA, 7, 689

\refb Surkova L. P., Svechnikov M. A. 2004, VizieR On-line Data Catalog: V/115

\refb Svechnikov M. A., Eretnova O. V., Olneva M. N., Taidakova T.
A. 1989, Nauchnye Informatsii, 67, 15

\end{document}